\documentclass[conference]{IEEEtran}
\pdfoutput=1
\IEEEoverridecommandlockouts
\usepackage{subfigure}
\usepackage{cite}
\usepackage{paralist}
\usepackage{multirow}
 \usepackage{siunitx}
\usepackage{amsmath,amssymb,amsfonts}
\usepackage{algorithmic}
\usepackage{graphicx}
\usepackage{textcomp}
\usepackage{xcolor}
\def\BibTeX{{\rm B\kern-.05em{\sc i\kern-.025em b}\kern-.08em
    T\kern-.1667em\lower.7ex\hbox{E}\kern-.125emX}}
\begin{document}

\title{C2S-AE: CSI to Sensing enabled by an Auto-Encoder-based Framework
}
\author{
\IEEEauthorblockN{Jun Jiang}
\IEEEauthorblockA{School of Communication and Information Engineering \\
Shanghai University, 200444 Shanghai, China \\
jun\_jiang@shu.edu.cn\\
}
~\\
\and
\IEEEauthorblockN{Shugong Xu*}
\IEEEauthorblockA{Xi'an Jiaotong-Liverpool University, 215123 Jiangsu, China \\
Shugong.Xu@xjtlu.edu.cn\\
*Corresponding author
}
~\\
\and
\IEEEauthorblockN{Wenjun Yu}
\IEEEauthorblockA{School of Communication and Information Engineering \\
Shanghai University, 200444 Shanghai, China \\
yuwenjun@shu.edu.cn \\
}
~\\
\and
\IEEEauthorblockN{Yuan Gao*}
\IEEEauthorblockA{School of Communication and Information Engineering \\
Shanghai University, 200444 Shanghai, China \\
gaoyuansie@shu.edu.cn \\
*Corresponding author
}

}

\maketitle

\begin{abstract}
Next-generation mobile networks are set to utilize integrated sensing and communication (ISAC) as a critical technology, providing significant support for sectors like the industrial Internet of Things (IIoT), extended reality (XR), and smart home applications. A key challenge in ISAC implementation is the extraction of sensing parameters from radio signals, a task that conventional methods struggle to achieve due to the complexity of acquiring sensing channel data. In this paper, we introduce a novel auto-encoder (AE)-based framework to acquire sensing information using channel state information (CSI). Specifically, our framework, termed C2S (CSI to sensing)-AE, learns the relationship between CSI and the delay power spectrum (DPS), from which the range information can be readily accessed. To validate our framework's performance, we conducted measurements of DPS and CSI in real-world scenarios and introduced the dataset 'SHU7'. Our extensive experiments demonstrate that the framework excels in C2S extrapolation, surpassing existing methods in terms of accuracy for both delay and signal strength of individual paths. This innovative approach holds the potential to greatly enhance sensing capabilities in future mobile networks, paving the way for more robust and versatile ISAC applications.
\end{abstract}


\section{Introduction}

In 2023, IMT-2030 outlined the vision for 6G, highlighting integrated sensing and communication (ISAC) as a key emerging use case\cite{recommendation2023framework}. Beyond communication, ISAC facilitates simultaneous sensing capabilities such as precise positioning, velocity estimation, and target detection\cite{gao2024performance}. These functionalities are expected to support a wide range of vertical industries, including the Industrial Internet of Things (IIoT), smart homes, and intelligent transportation systems\cite{liu2022survey}.

A crucial aspect of ISAC is parameter extraction, which converts radio echoes into vital information about the environment and objects\cite{zhang2022integration}. Solving these non-linear parameter estimation problems has led to the development of various algorithms, such as the periodogram (e.g., 2D discrete Fourier transform), subspace-based techniques like Multiple Signal Classification (MUSIC)\cite{wang2020angle}, Estimation of Signal Parameters via Rotational Invariant Techniques (ESPRIT)\cite{long2021aoa}, compressive sensing\cite{wu2022super}, and tensor-based methods\cite{zhang2021enabling}.

Despite these advancements, challenges remain in meeting the dual demands of sensing and communication in 6G scenarios: \begin{inparaenum}[\itshape a\upshape)]
\item Techniques like periodogram and subspace-based methods require substantial overhead, impacting communication performance\cite{zhang2021enabling};
\item Compressive sensing is computationally intensive and struggles with multiple target scenarios\cite{zhang2022integration};
\item Tensor tools rely on parameters estimated by other algorithms, inheriting their limitations.
\end{inparaenum}

In ISAC systems, valuable sensing information, e.g., ranging, can be extracted readily from the  delay power spectrum (DPS)\cite{ju2021millimeter}.  Conversely, channel state information (CSI) is more accessible due to established methods in mobile networks since 1G. Although a correlation between communication and sensing channels is intuitive, its exact nature remains unclear\cite{lou2023unified,zhang2023integrated}. The rise of artificial intelligence, particularly generative AI, presents opportunities to map CSI to DPS, potentially enhancing parameter extraction\cite{vaswani2017attention,cao2024survey}. By using the DPS, the range information can be readily accessed in both LoS and NLoS scenarios\cite{savic2014measurement}.

Our previous work pioneered this approach with a Transformer-based CSI to DPS\cite{gao2024c2s}. Despite demonstrating feasibility in practical networks, significant extrapolation errors were noted. To address this, we propose an auto-encoder (AE)-based framework to enable C2S (C2S-AE) in this paper, with the following key contributions:
\begin{itemize}
\item We present an AE-based framework that effectively learns the CSI-to-DPS correlation, enhancing both training and testing performance.
\item We introduce the 'SHU7' dataset, which includes DPS and CSI data from 94 Tx and Rx pairs measured at the Shanghai University station on metro line 7 in Shanghai, China. This dataset is used in this paper for model evaluation and will be publicly available for research reproducibility upon acceptance.
\item Using the 'SHU7' dataset, comprehensive simulation results demonstrate that the AE-based framework significantly outperforms the existing model in extrapolation accuracy, showing great potential to enable a more effective parameter extraction for ISAC systems.
\end{itemize}

The remainder of this paper is organized as follows: Section II discusses the problem of extrapolating DPS from CSI. Section III introduces our proposed C2S-AE framework. Section IV introduces the dataset 'SHU7' containing DPS and CSI, and the experiments of the proposed C2S-AE and existing model using this dataset. Finally, Section V concludes with a summary of our findings and future work.

\section{System model and problem formulation}

In wireless communication systems, understanding channel characteristics hinges on two fundamental concepts: CSI (Channel State Information) and DPS (Delay Power Spectrum). CSI provides a detailed overview of signal propagation between the transmitter and receiver, accounting for various channel effects. This information is particularly useful for techniques like precoding, which optimize signal transmission based on current channel conditions. For MIMO systems, CSI is denoted as the channel matrix \( \boldsymbol{H} \in \mathbb{C}^{L \times K} \) \cite{lu2014overview}, where \( L \) and \( K \) represent the number of receive and transmit antennas, respectively. The matrix \( \boldsymbol{H} \) is given by:
\begin{equation}
  \boldsymbol{H} = \left(
    \begin{array}{ccc}
      h_{11} & \cdots & h_{1K}\\
      \vdots & \ddots & \vdots\\
      h_{L1} & \cdots & h_{LK}\\
    \end{array}
  \right),
\end{equation}
where \( h_{ij} \) represents the channel coefficient between the \( i \)-th receive antenna and the \( j \)-th transmit antenna.

\begin{figure}[htbp]
\centering\includegraphics[width=0.5\textwidth]{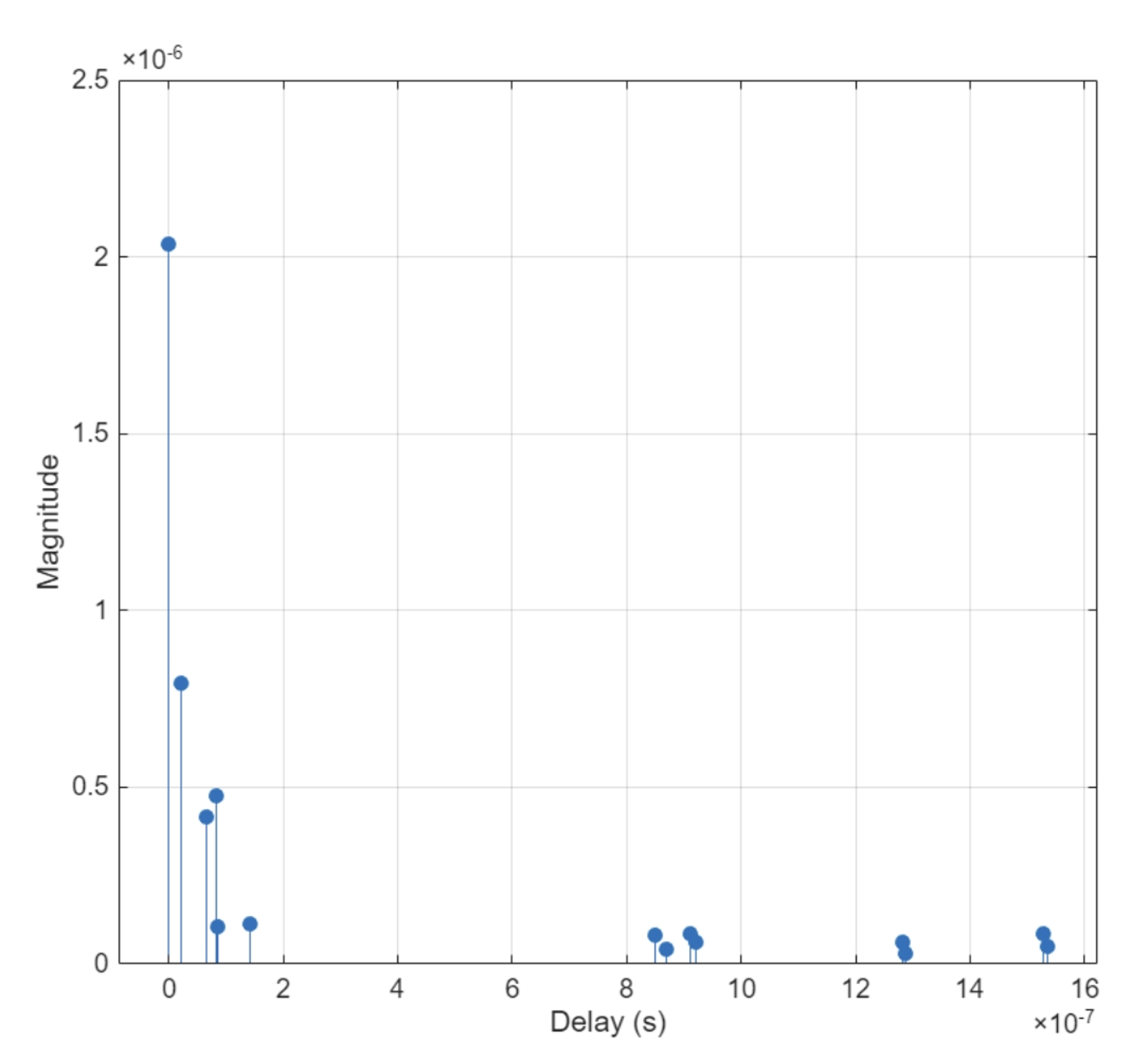}
\caption{An example of DPS in a multi-path scenario.}
\label{system_model_DPS}
\end{figure}

\begin{figure*}[htbp]
\centering\includegraphics[width=1\textwidth]{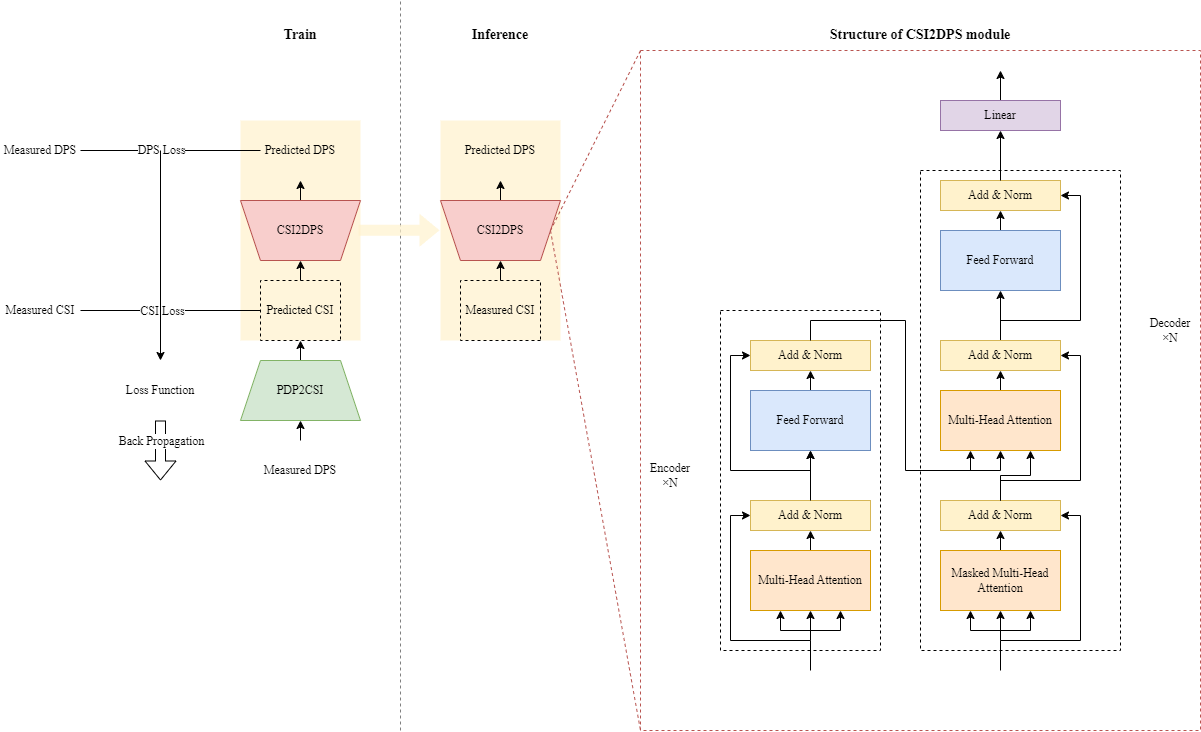}
\caption{Architecture of C2S-AE.}
\label{system_model}
\end{figure*}

In contrast, as illustrated in Fig. \ref{system_model_DPS}, DPS focuses specifically on multi-path propagation, offering detailed insights into the time delay and relative strength of signals traveling along different paths due to environmental factors such as reflections and scattering\cite{proakis2008digital,witrisal2001new}. This information is critical for understanding the temporal dispersion characteristics of the channel. 

While both CSI and DPS are vital for channel analysis, they serve distinct roles in system design and optimization. CSI's comprehensive perspective makes it suitable for overall channel assessment and advanced transmission techniques, whereas DPS's focus on multi-path effects is essential for analyzing signal delays and power distribution in complex environments. By using the DPS, the range information can be readily accessed in both LoS and NLoS scenarios\cite{savic2014measurement}.

Traditional methods for measuring DPS typically require intricate time-domain measurements at the receiver, involving expensive hardware and considerable time investment, especially in complex environments. However, extrapolating DPS data from CSI can significantly reduce measurement costs, making channel characteristic analysis more convenient and efficient.

Moreover, in rapidly changing communication environments, where channel states evolve dynamically, real-time acquisition and updating of DPS information are crucial for adaptive optimization of communication systems. By leveraging the latest CSI data to extrapolate DPS, the system can swiftly respond to channel changes, adjust transmission strategies, and ensure communication stability and efficiency.

\section{Proposed framework}
As illustrated in Fig. \ref{system_model}, we propose an AE-based framework to extrapolate the DPS using CSI. AE models generally leverage unsupervised learning to derive a compact representation from input data through two main components: the encoder and the decoder. The encoder compresses the input data into a lower-dimensional representation, while the decoder reconstructs the original data from this compressed form.

However, we propose an improved AE framework that adopts a supervised learning approach, thereby enhancing the performance of the model to extrapolate DPS. Specifically, the output of the encoding is designed to be CSI, while the decoder reconstructs the predicted CSI back into DPS.

The improved framework introduces two critical enhancements: First, by explicitly modeling CSI as the latent variable of DPS, the latent representation acquires well-defined physical significance, which facilitates system debugging and theoretical analysis. Second, the proposed joint loss function (combining CSI reconstruction and DPS reconstruction) simultaneously constrains both the latent space and output space, enabling better capture of the physical correlations between DPS and CSI compared with end-to-end models employing single-target optimization (DPS-only). This dual-constraint mechanism ensures the learned latent space preserves essential physical characteristics while maintaining high-fidelity DPS reconstruction capability.

\subsection{Encoder}

Let $\mathbf{P}$ represent the DPS and $\mathbf{C}$ the corresponding CSI. The encoder is defined as a nonlinear mapping $f_E(\mathbf{P}; \theta_E)$, where $\theta_E$ are the parameters of the encoder. This mapping converts the DPS $\mathbf{P}$ into a lower-dimensional vector $\mathbf{z} = f_E(\mathbf{P}; \theta_E)$. Ideally, this representation $\mathbf{z}$ should capture essential features of the input and closely resemble the target CSI $\mathbf{C}$, i.e., $\mathbf{z} \approx \mathbf{C}$.

\subsection{Decoder}
The decoder reconstructs the original DPS $\mathbf{P}$ using the output $\mathbf{z}$ of the encoder. Similarly, the process is denoted by a nonlinear mapping $f_D(\mathbf{z}; \theta_D)$, where $\theta_D$ are the parameters of the decoder. The reconstructed DPS $\hat{\mathbf{P}}$ should be nearly identical to the original input DPS $\mathbf{P}$, expressed as:

\begin{equation}
\hat{\mathbf{P}} = f_D(f_E(\mathbf{P}; \theta_E); \theta_D) \approx \mathbf{P}
\end{equation}
\begin{figure*}[htbp]\centering    
\includegraphics[width=1.65\columnwidth]{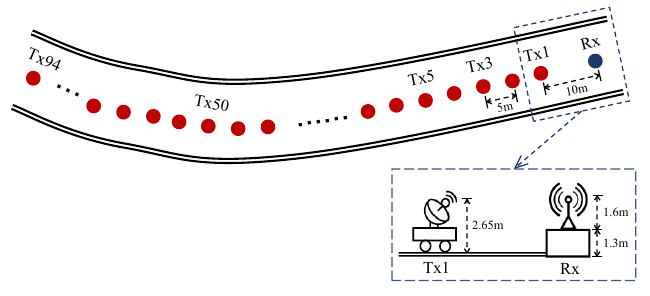} 
\caption{The layout and Tx \& Rx position of channel measurement.}   
\label{measurement_setup}\end{figure*}

\subsection{Training Process}

Given a training set $\{(\mathbf{P}_i, \mathbf{C}_i)\}_{i=1}^N$, where each sample pair $(\mathbf{P}_i, \mathbf{C}_i)$ includes a DPS $\mathbf{P}_i$ and its corresponding CSI $\mathbf{C}_i$. $\mathbf{P}_i \in \mathbb{R}^{N_p \times 1023}$, with $N_p$ denoting the number of measurement points; and $1023$ representing the number of time delay points. $\mathbf{C}_i \in \mathbb{R}^{N_p \times 2}$, with $N_p$ indicating the number of measurement points; and $2$ representing the magnitude and phase of the CSI.

Both the encoder and decoder employ a standard Transformer architecture with six layers, excluding positional encoding. This omission is due to the inherent structure of the input data, which already contains sufficient positional information. The Transformer's architecture allows for predicting DPS at any continuous $N_p$ measurement points from CSI at any continuous $N_p$ measurement points, enhancing the flexibility and applicability of the model.

During training, each DPS sample $\mathbf{P}_i$ is fed into the encoder to produce an intermediate representation $\mathbf{z}_i$, which is then used by the decoder to reconstruct $\mathbf{P}_i$. The training objective is to minimize both the reconstruction error and the difference between $\mathbf{z}_i$ and the true CSI $\mathbf{C}_i$.
\begin{align}
&L(\theta_E, \theta_D)=\\ & \frac{1}{N} \sum_{i=1}^{N} \left( \left\| \mathbf{P}_i - f_D(f_E(\mathbf{P}_i; \theta_E); \theta_D) \right\|^2 + \left\| \mathbf{C}_i - f_E(\mathbf{P}_i; \theta_E) \right\|^2 \right)    
\end{align}

This approach enables the encoder to learn effective CSI representations from DPS and allows the decoder to accurately reconstruct the original DPS from these representations.

\section{Experiment}
\subsection{'SHU7' Measurement}

In this work, we introduce the 'SHU7' dataset, which contains CSI and DPS collected at a 3.5 GHz frequency band in an indoor environment. As shown in Fig. \ref{measurement_setup}, the measurement campaign was conducted using 64 transmitter (Tx) antennas and 32 receiver (Rx) antennas, mounted on a platform trolley at the Shanghai University Line 7 subway station. Spread spectrum communication was utilized to enhance the signal-to-noise ratio at the receiver. Electrical switches controlled the Tx and Rx antennas to simulate a $64 \times 32$ massive MIMO channel. The transmission bandwidth was 160 MHz with a PN10 channel type, while the receiver's sampling rate was 200 MHz, with 128 PN sequence periods.

\begin{figure}[htbp]
\centering\includegraphics[width=1\columnwidth]{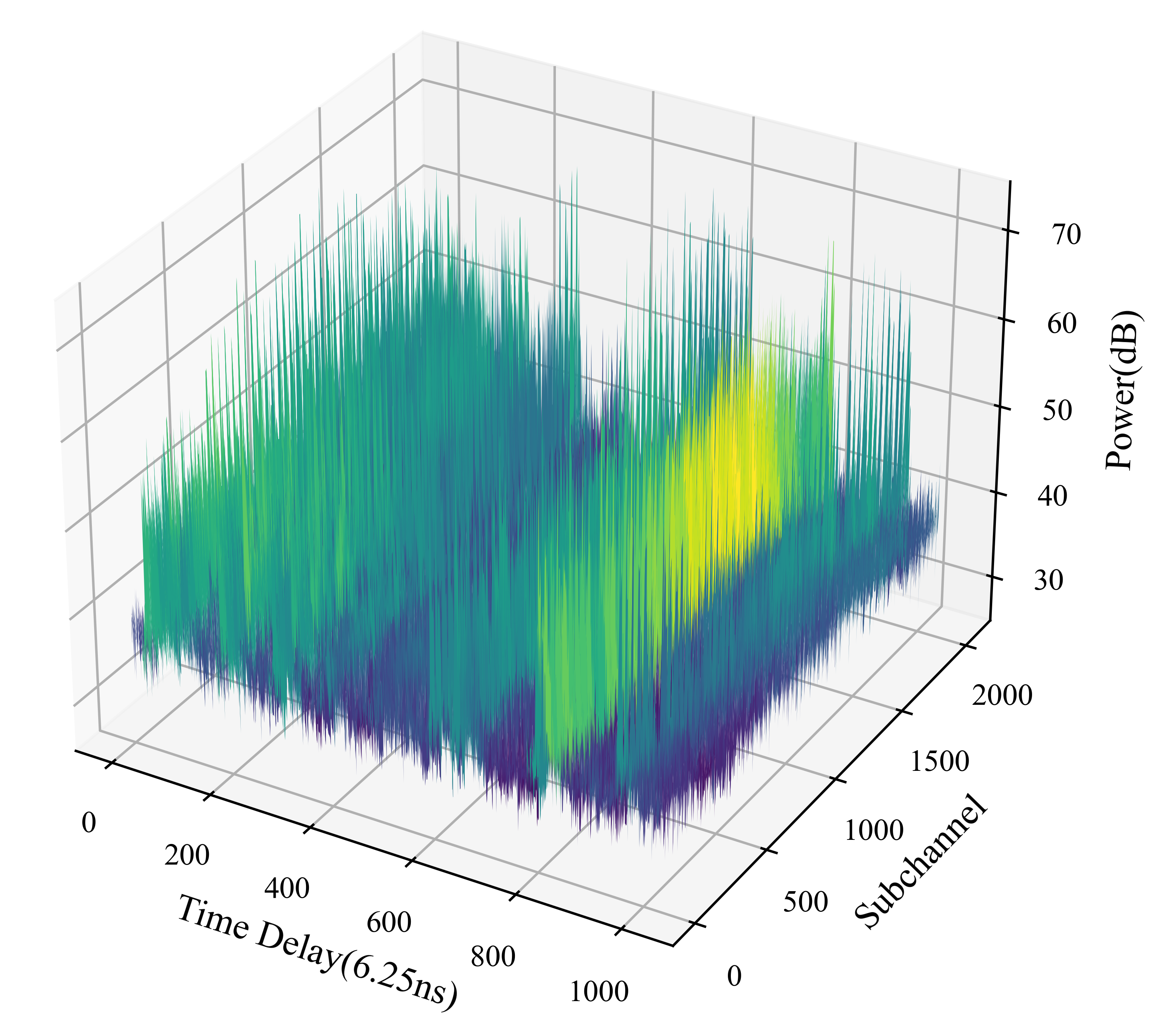}
\caption{Illustration of measured DPS.}
\label{DPS_dia}
\end{figure}

Initially, the distance between the Tx and Rx was 10 meters, and the Rx antenna was moved 5 meters for each subsequent measurement, eventually reaching a distance of 400 meters. At each measurement point, a channel impulse response (CIR) was recorded every 40 microseconds. These measurements yielded CIR samples denoted as $h_m \in \mathbb{R}^{128 \times 1023 \times 2048}$, where 128, 1023, and 2048 represent the number of time periods, delay points, and sub-carriers, respectively. We squared the amplitude of each CIR to obtain the corresponding DPS, as illustrated in Fig. \ref{DPS_dia}, and performed a Fourier transform to sample at the center frequency of each CIR to derive the CSI. The resulting dataset comprises 192,512 samples, each containing both CSI and DPS data, with 100,352 Line-of-Sight (LoS) samples and 92,160 Non-Line-of-Sight (NLoS) samples.

\begin{figure*}[htbp]
\centering\includegraphics[width=0.8\textwidth]{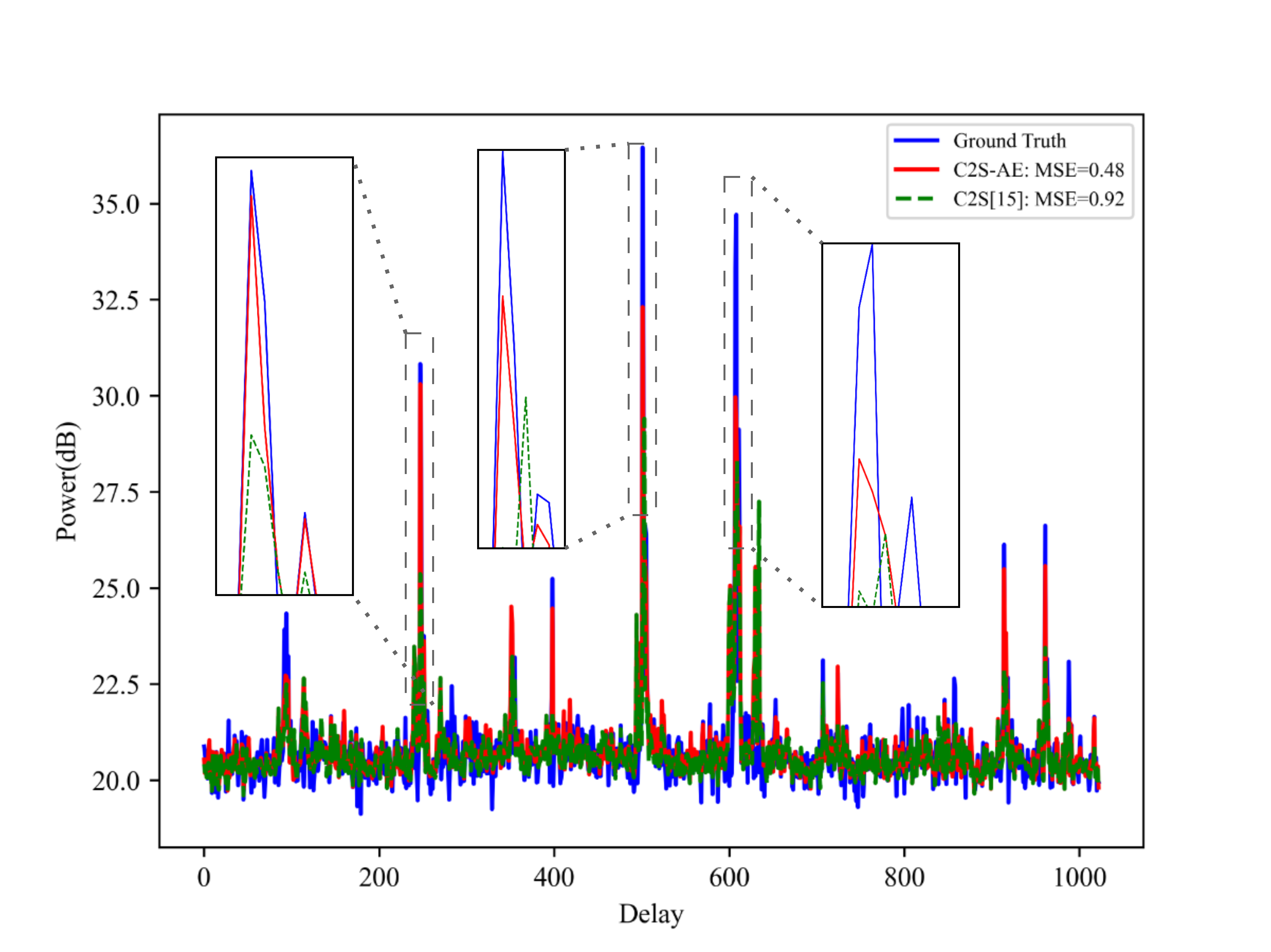}
\caption{Extrapolated DPS using C2S\cite{gao2024c2s} and proposed C2S-AE.}
\label{visual}
\end{figure*}

\begin{table*}[tbh]\centering
\label{tab:results}
\begin{tabular}{cc|cccccc}\hline
 \multicolumn{2}{c|}{$N_p$} & 1 & 2 & 4 & 8 & 16 & 32 \\ \hline
\multicolumn{1}{c|}{\multirow{2}{*}{MSE}} & C2S\cite{gao2024c2s} & 0.885 & 0.885 & 0.846 & 0.786 & 0.657 & 0.567  \\
\multicolumn{1}{c|}{} & Proposed C2S-AE & 0.759 & 0.749 & 0.728 & 0.711 & 0.604 & 0.521  \\
\multicolumn{1}{c|}{} & Improvement & \textbf{14.2\%} & \textbf{12.4\%} & \textbf{13.9\%} & \textbf{9.54\%} & \textbf{8.07\%} & \textbf{8.11\%}  \\\hline
\multicolumn{2}{c|}{Inference Time(ms)} & 4.91 & 5.57 & 5.90 & 6.48 & 7.29 & 7.68 \\\hline
\end{tabular}
\caption{Comparison between C2S\cite{gao2024c2s} and proposed C2S-AE}
\end{table*}
\subsection{Results Analysis}

While CSI can be directly derived from DPS through traditional computations and used for prediction via deep learning methods by training only the decoder of the AE\cite{gao2024c2s}, our experimental results demonstrate that using a complete AE architecture significantly enhances predictive accuracy, as shown in Fig. \ref{visual}. This improvement is attributed to the challenge of mapping from a 2-dimensional space to a 1023-dimensional space when only training the decoder. The inclusion of an encoder, which reduces dimensionality from 1023 to 2, aids the decoder in learning the mapping from CSI to DPS, thereby optimizing overall system performance.

Table I compares the Mean Squared Error (MSE) performance of the proposed autoencoder against a model trained solely on the decoder across different numbers of neighboring points $N_p$. 

Consider that training only the decoder forces it to map from an arbitrary 2D CSI space to the complex 1023D DPS, which is akin to solving an underdetermined problem. However, C2S-AE, learned end-to-end, captures critical dependencies between CSI and DPS, improving predictive accuracy. The encoder learns to prioritize features critical for the decoder's task, creating a synergistic relationship. In contrast, a decoder trained in isolation lacks this feedback loop, leading to suboptimal. C2S-AE reduces the complexity of the mapping task, captures essential CSI-DPS relationships, and enables joint optimization, resulting in 8.07\% lower MSE and superior prediction accuracy.

However, as $N_p$ increases, the MSE gap between the two models narrows. This may be because the model trained only on the decoder begins to effectively utilize additional information with larger $N_p$, enhancing its predictive performance. Nevertheless, it still cannot match the performance of our proposed method.

Furthermore, Table I indicates that inference time increases with more neighboring points $N_p$. Specifically, as $N_p$ rises from 1 to 32, inference time grows from 4.91 ms to 7.68 ms, reflecting the greater computational demand of processing more information. These times were measured on an NVIDIA GeForce RTX 4090 GPU with a batch size of 1, averaged over 1000 repetitions.

\section{conclusions}
Extrapolating the DPS using CSI presents a promising method for parameter extraction in ISAC systems. However, existing Transformer-based frameworks have shown limited performance in this area. In this paper, we introduce the C2S-AE framework, an innovative AE-based framework designed to effectively learn the correlation between the sensing channel, i.e., DPS, and the communication channel, i.e., CSI, and further extrapolate DPS using CSI. To evaluate the performance of proposed C2S-AE, We introduced the 'SHU7' dataset, which will be open to public later, containing DPS and CSI data measured from 94 Tx and Rx pairs measured at the Shanghai University station on metro line 7 in Shanghai, China. Using this dataset, extensive experiment demonstrates that the proposed C2S-AE framework outperforms the existing framework in term of extrapolation accuracy for both the delay and signal strength of individual paths. This innovative approach holds the potential to greatly enhance sensing capabilities in future mobile networks, paving the way for more robust and versatile ISAC applications.

Building on these promising results, future work will focus on applying the DPS extrapolated from CSI to wireless localization and other sensing tasks. The precise estimation of DPS opens new opportunities for localization and tracking in complex environments, which can significantly enhance both indoor and outdoor positioning systems. This will enable more accurate user and device localization for applications such as autonomous navigation and smart cities. Moreover, the C2S-AE framework can be further optimized to improve its scalability and robustness in real-world scenarios, broadening its applicability in diverse use cases.

\subsection*{Acknowledgments}
The authors would like to thank Prof. Guoxin Zheng of Shanghai University, China, for leading and management the CSI and DPS measurement at the Shanghai University station on metro line 7 in Shanghai, China.
\bibliographystyle{IEEEtran}
\bibliography{bibfile}
\end{document}